\input{aipcheck}

\documentclass[
    ,final            
  ]
  {aipproc}

\layoutstyle{6x9}

\begin{document}

\title{VLBI observations of Brightest Cluster Galaxies: are cooling and parsec scale morphology correlated?}


\author{E. Liuzzo}{
  address={Istituto di Radioastronomia, INAF, via Gobetti 101, 40129 Bologna, Italy},
  email={liuzzo@ira.inaf.it},
  altaddress={Dipartimento di Astronomia, Universit\`a di Bologna, via Ranzani 1, 40127 Bologna, Italy
}
}
\author{G. Giovannini}{
  address={Istituto di Radioastronomia, INAF, via Gobetti 101, 40129 Bologna, Italy},
  email={ggiovann@ira.inaf.it},
  altaddress={Dipartimento di Astronomia, Universit\`a di Bologna, via Ranzani 1, 40127 Bologna, Italy
}
}
\author{M. Giroletti}{
  address={Istituto di Radioastronomia, INAF, via Gobetti 101, 40129 Bologna, Italy},
  email={giroletti@ira.inaf.it},
}
\begin{abstract}
 We present a statistical study on parsec scale properties of a sample of Brigthest Cluster Galaxies (BCGs) in Abell Clusters. These data show a possible difference between BCGs in cool core clusters (two-sided parsec scale jets) and in non cool core clusters (one-sided parsec scale jet).
We suggest that the two-sided morphology in cool core clusters could be due to the presence of
mildly relativistic jets slowed down already at mas scale as consequence of the jet interaction with a dense surrounding medium.

\end{abstract}

\maketitle


\section{Introduction.}

Brightest Cluster Galaxies (BCGs) are a unique class of objects being the most luminous and massive galaxies in the Universe \citep{ye04} . 

In the radio band, BCGs are more likely to be radio-loud than other galaxies of
the same mass \citep{be06}.  Some BCGs have a standard tailed structure,
either extended on the kpc scale (e.g. 3C465 in A2634), or with a small size
(e.g. NGC4874 in Coma cluster).  In other cases, they show a diffuse and
amorphous radio morphology that is rare in the general radio population, but
very often present in BCGs in cooling core clusters of galaxies. The presence of
X-ray cavities in the emitting gas coincident with the presence of radio lobes
can also be the proof of the interplay between the radio activity of BCGs and
the arrest or slow down of the cooling process at the cluster centers \citep{bi08, du06}. This scenario is in agreement with the recent result that in every/most of cooling core clusters is present an active radio BCG \citep{ei02}.

However, a few points need a deeper study: not all BCGs are strong radio
sources and a restarting activity with a not too long duty cycle is necessary
to justify the slow down of cooling processes.  Moreover, it is not clear if radio properties of
BCGs in cooling flow clusters are systematically different from those of BCGs
in merger clusters; we note that in cooling clusters the kpc scale radio morphology
of BCGs is diffuse and relaxed and it is often surrounded by an extended low-brightness radio emission that takes the form of mini-halo \citep{gi07}. However, extended $`$normal' sources have been also found (e.g. Hydra A, [9]). In merging clusters the most common morphology is a Wide Angle Tail source but point-like as well as core-halo sources are also present.  

On the parsec scale, BCGs are not yet well studied as a class of sources.  Only
a few of them have been observed, being well known radio galaxies.  In some
cases, they look like normal FRI radio galaxies with relativistic collimated
jets. Jets are often one-sided because of Doppler boosting effects (e.g.  3C465
in A2634 and 0836+29 in A690 \citep{ven95}), although there are also
cases where two-sided symmetric jets are present in VLBI images, and the
presence of highly relativistic jets is not certain (e.g. 3C338 in A2199
\citep{ge07}, and Hydra A in A780 \citep{tay96}).
\section{The project: observations and results.}

In order to realize a statistical study of parsec scale properties of BCGs we selected a complete sample of BCGs in nearby Abell clusters with the following constraints: 1) Distance Class $\leq$2, and  2) Declination $> 0^\circ$.
All clusters have been included without constraints on cluster conditions (e.g cooling) and no 
selection is present on the BCG radio power. In the complete sample, we have 27 BCGs, including cases like A400 (double BCG, 3C75, Fig. 1) and clusters with a clear double structure (e.g. A1314) where we observed the BCG of both substructures. 

\begin{figure}
\centering
\includegraphics[width=0.28\textwidth]{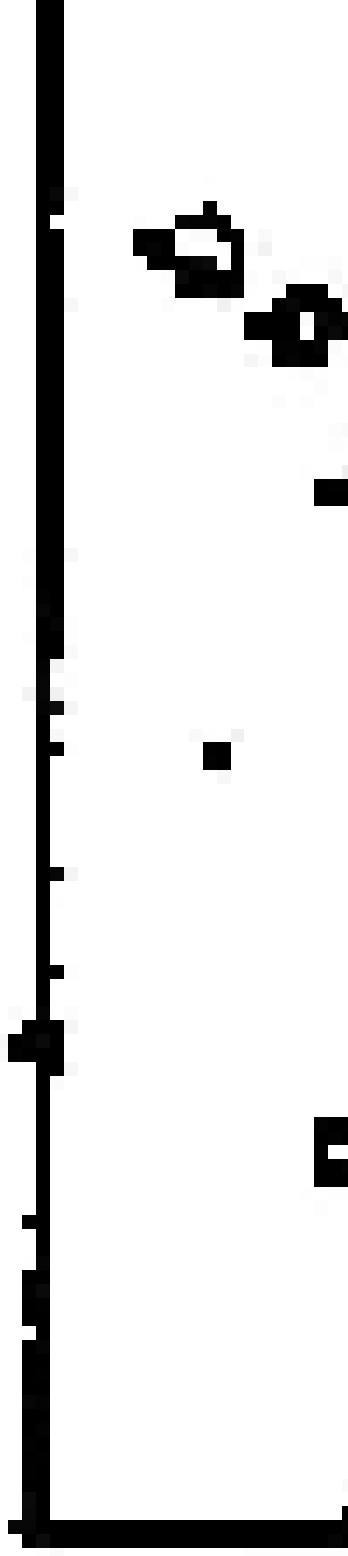}
\hfill
\includegraphics[width=0.28\textwidth]{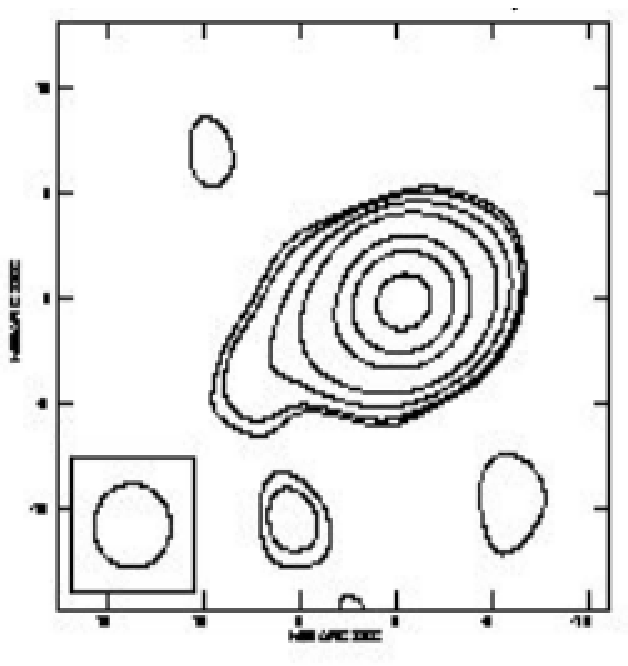}
\hfill
\includegraphics[width=0.3\textwidth]{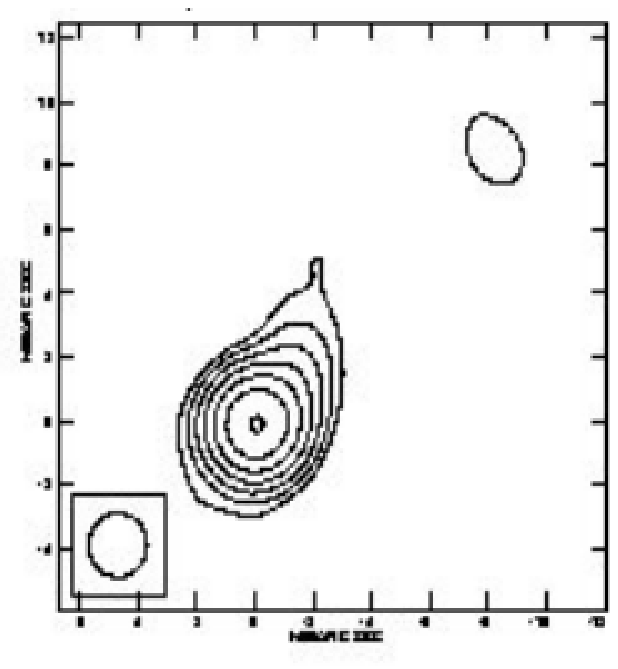}
\caption{On the left, VLA image \citep{ei02} at 6 cm of 3C 75A and 3C 75B, the double BCG in the non cool core cluster A400. Contour intervals are (-1, 1, 2, 4, 8, 16, 32, 64, 128, 256, 512) $\times$ 0.1 mJy/beam. The HPBW is 1.4 $\times$ 1.4 arcsec with P.A. = 0$^{\circ}$. At the center: one-sided VLBI image \citep{liu09} of 3C75A. Contours levels are 0.3, 0.6, 1.2, 2.4, 4.8, 9.6, 19.2 mJy per clean beam and the HPBW is 2 $\times$ 2 mas with P.A. = 0$^{\circ}$. On the right: one-sided VLBI image \citep{liu09} of 3C75B. Contours levels are -0.3, 0.3, 0.4, 0.8, 1.8, 7, 14, 28 mJy/beam and the HPBW is 4 $\times$ 4 mas with P.A. = 0 $^{\circ}$.}
\end{figure}

We collected for all 27 sources of the complete sample VLBA observations at 5 GHz in phase referencing mode that allow typically resolution of 3 $\times$ 1.8 mas and noise level in our final maps of 0.1 mJy/beam. 
To improve our statistics, we
performed a search in the literature and archive data looking for VLBI
data of BCGs in Abell clusters with DC $>$2. We added to our complete sample the following clusters: A690, A780, A1795, A2052, A2390, and A2597 (expanded sample).

Among the sources of the extended sample, we analized in particular 4C 26.42, the BCG of A1795 (Fig. 2). In \citep{liu09} we presented our results: the interaction with the surrounding medium seems to be fundamental in order to understand the peculiar radio morphology of this BCG.

Statistical results are presented in Table 1. In the expanded sample, we find a remarkable dominance of two-sided sources in relaxed clusters (70\%), and of one-sided  (56\%) or non-detected (39\%) sources in merging systems. No one-sided source is present in cool core clusters, and no two-sided source is present in non cool core clusters. We have to note that most of undetected sources in VLBA observations are in BCG that are radio quiet (or faint) also in VLA observations.

\begin{table}
\begin{tabular}{ccccccc}
\hline
\hline
Sample & Cluster  & Number & two-sided & one-sided & point & N.D. \\
       & morphology &        &           &           &       &      \\
\hline 
Complete & cool core    &  5     &  2 (40$\%$)       &   --      & 1     & 2    \\
         & non cool core    &  22    & --        &  12 (55$\%$)      & 1     & 9    \\
\hline
Expanded & cool core   &  10    & 7 (70$\%$)   & --  & 1  & 2  \\
         & non cool core  &  23    & --  & 13 (56$\%$) & 1  & 9  \\
\hline
\end{tabular}
\caption{{\bf BCG counts in the complete (nearby) sample and expanded one}.  We
  report the number of BCG according to the cluster morphology and pc scale
  morphology. }
\end{table}

The difference in the parsec scale jet structure between the relaxed and non relaxed clusters is evident. We used as a comparison sample the Bologna Complete Sample (BCS) \citep{gio01, gio05}. It is a complete sample characterized by no selection effect on jet velocity and orientation. In this sample, 73\% of FRI radio galaxies (in the same radio power range of our sample of BCGs) have a one-sided parsec scale jet structure because of relativistic effects. This percentage corresponds to a random angle ditribution of sources with relativistic jets. From the comparision, we concluded that FRI sources not BCG and BCG in merging clusters have similar parsec scale properties, BCGs in cool core clusters show different parsec scale properties. 

Due to the statistical results for the BCS, we conclude that one sided parsec scale structures in BCGs in merging clusters are in agreement with relativistic effects in intrinsically symmetric jets. Instead, in relaxed clusters, two-sided jets could the consequence of relativistic jets on the plane of the sky or of mildly relativistic jets randomly oriented. The first hypothesis is excluded for statistical considerations. The percentage of two-sided parsec scale structures in the sample of BCGs is too high  (70\%) compared this (23\%) for the FRI radio galaxies not BCGs of the BCS. Therefore, the second suggestion seems to be the most reasonable. \citep{ro08} showed that a jet perturbation grows because of Kelvin- Helmotz instability and produce a strong interaction between the jet and ISM with a consequent jet decelaration. The decelaration is more efficient if the density ratio between of the ISM and the jet is high as expected at the center of cooling clusters \citep{sal03}. In BCGs in cooling clusters relativistic jets could slow down on the parsec (sub-pc) scale. For a comparison, we note that in FRI relativistic jets slow down on the sub-kpc scale.  
\begin{figure}
  \includegraphics[height=.4\textheight]{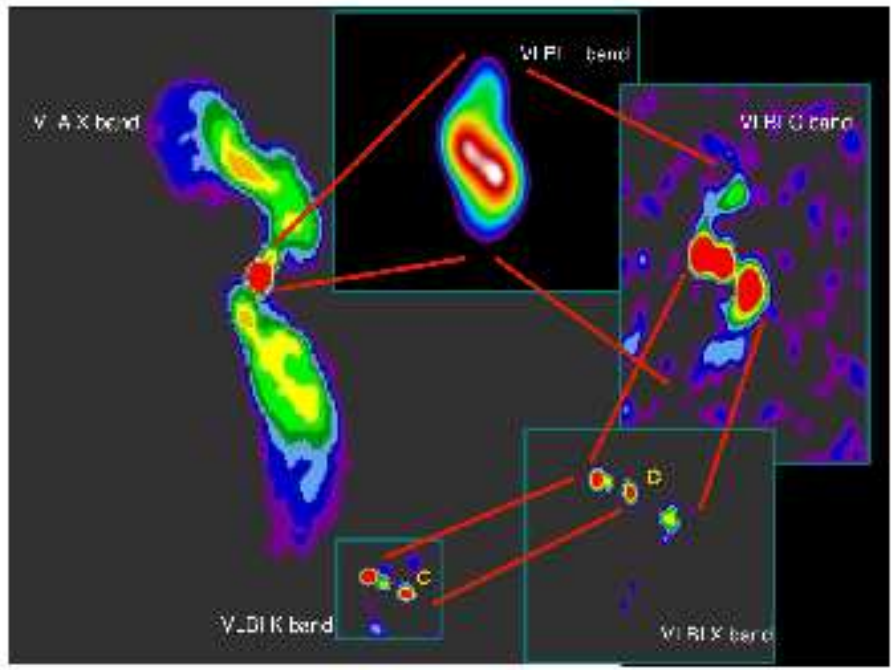}
  \caption{Clockwise from left to right, zooming from kiloparsec to mas scale radiostructure of 4C 26.42: color maps of VLA X band, VLBI L band, VLBI C band, VLBI X band and VLBI K band data. (C) indicates the core component. }
\end{figure}

We conclude that the possibile dichotomy between BCGs in cool core and non cool core clusters could be due not to intrinsic jet differences but to different ISM conditions. We plan to observe a larger sample of BGCs in cooling and relaxed clusters with VLBA to improve our statistic.

\begin{theacknowledgments}
  We thank the organizers of a very interesting meeting. The National Radio Astronomy Observatory is operated by Associeted Universities, Inc., under cooperative agreement with the National Science Foundation.

\end{theacknowledgments}






\end{document}